\documentclass[pra,aps,eqsecnum,amsmath,amssymb,showkeys,showpacs]{revtex4}
\usepackage{graphicx}

\newcommand{\D}{\mathsf{D}}
\newcommand{\T}{\mathsf{T}}
\begin{document}
\clearpage
\preprint{}

\title{Optimal cloning with respect to the relative error}
\author{Alexey E. Rastegin}
\affiliation{Department of Theoretical Physics, Irkutsk State University,
Gagarin Bv. 20, Irkutsk 664003, Russia}

\begin{abstract}
The relative error of cloning of quantum states with arbitrary prior
probabilities is considered. It is assumed that the ancilla may
contain some {\it a priori} information about the input state to be cloned.
The lower bound on the relative error for general cloning scenario
is derived. Both the case of two-state set and case of multi-state
set are analyzed in details. The treated figure of merit is compared
with other optimality criteria. The quantum circuit for optimal
cloning of a pair of pure states is constructed.
\end{abstract}
\pacs{03.65.ta, 03.67-a}
\keywords{State-dependent quantum cloning; stronger no-cloning
theorem; distinguishability transfer gate.}

\maketitle

\section{Introduction}

The quantum information topics are the subject of active
research \cite{hayashi}. The impressive progress have been reached
in the quantum cryptography \cite{gisin} and study of quantum
algorithms \cite{vandam}. Due to an impact on security in quantum
cryptography, the quantum cloning is still a significant topic. At
the same time, a cloning itself is hardly sufficient for an
eavesdropping \cite{palma,pirandola}. No-copying results have been
established for pure states \cite{wootters,dieks} as well as for
mixed states \cite{barnum}. In view of such evidences, the question
arose how well quantum cloning machines could work. In effect, the
basic importance of the no-cloning theorem is expressed much
better in more detailed results, which also give explicit bounds
on an amount of the noise.

After the seminal work by Bu\v{z}ek and Hillery \cite{hillery1},
many approaches to approximate quantum cloning have been
developed. In view of existing reviews \cite{fiurasek,fanh}, we
cite only the literature that is directly connected to our
results. An approximate cloning of two prescribed pure states was
first considered in Ref. \cite{hillery2}. This kind of cloning
operation is usually referred to as {\it state-dependent
cloning} \cite{bruss1}. In general, various types of
state-dependent cloners may be needed with respect to the question
of interest \cite{pati07,siomau}. Errors inevitably occur already
in a cloning of two nonorthogonal states \cite{hillery2}. How close
to perfection can a cloning be? Of course, any explicit answer
must utilize some optimality criterion. We will refer criterion
used in Ref. \cite{hillery2} to as the {\it absolute
error} \cite{rast1}. Chefles and Barnett \cite{chefles2} derived the
least upper bound on the global fidelity for cloning of two pure
states with arbitrary prior probabilities. The quantum circuit
that reaches this upper bound was also constructed \cite{chefles2}.
The global fidelity of cloning of several equiprobable pure states
was examined in Ref. \cite{guo}.

Although cloning problems were mostly analyzed with respect to the
fidelity criteria, other measures of closeness of quantum states
are relevant. For example, the ''partial'' quantum cloning is
easier to analyze with respect to the squared Hilbert-Schmidt
distance \cite{kazakov}. One of criteria, {\it relative
error} \cite{rast1}, has been shown to be useful within the B92
protocol emerged in Ref. \cite{bennett3}. Deriving bounds on
the relative error was based on the spherical triangle
inequality\cite{rast1} and the notion of the angle \cite{rast2} sometimes called
the {\it Bures length} \cite{uhlmann09}. Using this new method, a cloning of two
equiprobable mixed states was studied with respect to the global
fidelity \cite{rast2}. The results of Ref. \cite{rast1} were
partially extended to mixed-state cloning \cite{rast3}.

In a traditional approach, the ancilla does not contain {\it a
priori} information of state to be cloned just now. A more general
case is the scope of the stronger no-cloning theorem \cite{jozsa1}.
Namely, a perfect cloning is achievable, if and only if the full
information of the clone has already been provided in the ancilla
state alone. In Ref. \cite{rast4} we examined a cloning of
finite set of states when the ancilla contains a partial
information of the input state. So, the previous result of
Ref. \cite{chefles2} was extended to both the mixed states and
{\it a priori} information.

In this paper, we study the relative error of cloning of several
mixed states, having arbitrary prior probabilities. {\it A priori}
information in the ancilla is also assumed. In Section II, the
relative error criterion introduced in Refs. \cite{rast1,rast3}
is extended to the general cloning scenario. We derive the lower
bounds on the relative error for cloning of two-state set (see
Section III) and multi-state set (see Section IV). In Section V,
the relative error is compared with other optimality criteria. We
also build the quantum circuit for cloning of two pure states (see
Section VI). This circuit reaches the lower bound on the relative
error for arbitrary prior probabilities and {\it a priori}
knowledge about the input. Section VII concludes the paper.

\section{Relative error of cloning}

The main problem posed formally is this. We have $N$
indistinguishable $n$-level systems that which are all prepared in
the same state $\rho_j\in\{\rho_1,\ldots,\rho_m\}$ from the known
set of density operators on the space
${\cal{V}}={\mathbb{C}}^{n}$. These $N$ systems form the register
$B$. Its initial state is a density operator
$\varrho_j=\rho_j^{\otimes N}$ on the input Hilbert space
${\cal{H}}\equiv{\cal{V}}^{\otimes N}$. The prior probabilities
$p_j$ of states $\varrho_j$ obey the normalization condition
$\sum\nolimits_{j=1}^{m} p_j=1$. We aim to get a larger number $L$
of copies of the given $N$ originals by means of the ancilla whose
initial state is $\Upsilon_j$ according to the input $\varrho_j$.
Here we mean a system $CE$ composed of extra register $C$ and
environment $E$. The extra register $C$ contains $M=L-N$
additional $n$-level systems, each is to receive the clone of
$\rho_j$. If we include an environment space then any
deterministic physical operation may be expressed as a unitary
evolution. Thus, the final state of two registers $BC$ is the
partial trace over environment space
\begin{equation}
\varrho'_j={\mathcal{E}}(\varrho_j)={\rm Tr}_{E}
\Bigl[{\mathsf{U}}(\varrho_j\otimes\Upsilon_j)
{\mathsf{U}}^{\dagger}\Bigr] \ .
\label{bd2}
\end{equation}
The output $\varrho'_j$ is a density operator on the output
Hilbert space ${\cal{H}}'\equiv{\cal{V}}^{\otimes L}$.

The actual output $\varrho'_j$ must be compared with the ideal
output $\rho_j^{\otimes L}$. Many measures of distinguishability between
mixed quantum states are based on the fidelity \cite{maz}. We shall employ the angles
and the sine metric \cite{rast06}. Let $|{\mathsf{A}}|$ denote a
unique positive square root of
${\mathsf{A}}^{\dagger}{\mathsf{A}}$. The fidelity  between the two density operators
$\sigma$ and $\omega$ is equal to
$F(\omega,\sigma)=\bigl({\rm{Tr}}|\sqrt{\omega}\sqrt{\sigma}|\bigr)^2$ \cite{uhlmann1,jozsa2}.
In terms of this measure, the angle
$\Delta(\omega,\sigma)\in[0;\pi/2]$ between $\sigma$ and $\omega$
is defined by the equality
$\cos^2\Delta(\omega,\sigma)=F(\omega,\sigma)$ \cite{rast2}. It is also
referred to as the {\it Bures length} \cite{uhlmann09}, because of
its close relation to the standard Bures metric
$d_B(\omega,\sigma)=\bigl(2-2\sqrt{F}(\omega,\sigma)\bigr)^{1/2}$. Due
to the spherical triangle inequality \cite{rast2},
\begin{equation}
\Delta(\omega,\sigma)\leq
\Delta(\omega,\eta)+\Delta(\sigma,\eta)
\ . \label{pr10}
\end{equation}
We introduce the {\it sine distance} \cite{rast06} between $\omega$
and $\sigma$ as $d(\omega,\sigma):=\sin\Delta(\omega,\sigma)$.
This metric on the space of quantum states has a close relation to
the trace distance \cite{rast07} and enjoys the
following \cite{rast3,rast06}. For any POVM measurement
$\{{\mathsf{A}}_{\mu}\}$, there holds
\begin{equation}
\bigl|{\rm{Tr}}({\mathsf{A}}_{\mu}\omega) -
{\rm{Tr}}({\mathsf{A}}_{\mu}\sigma) \bigr| \leq
d(\omega,\sigma) \ .
\label{pr11}
\end{equation}
Here ${\rm{Tr}}({\mathsf{A}}_{\mu}\omega)$ is the probability of
obtaining outcome $\mu$, if the state right before measurement was
$\omega$. A more detailed characterization of such a kind can be
posed via majorization relations \cite{rast091}. We also have
$|F(\omega,\eta)-F(\sigma,\eta)|\leq d(\omega,\sigma)$. Since the
fidelity function cannot decrease under any deterministic quantum
operation \cite{barnum}, the last inequality can be extended to
\begin{equation}
\bigl| F({\mathcal{E}}(\omega),\eta') -
F({\mathcal{E}}(\sigma),\eta') \bigr| \leq
d(\omega,\sigma) \ .
\label{pr14}
\end{equation}
Using the sine distance is reasonable approach due to the
inequalities (\ref{pr11}) and (\ref{pr14}). For brevity, let us
denote
\begin{equation}
\delta'_j \equiv \Delta(\varrho'_j,\rho_j^{\otimes L})
\ , \qquad \Delta_{jk}^{(L)} \equiv
\Delta(\rho_{j}^{\otimes L},\rho_{k}^{\otimes L}) \ . \label{bd4}
\end{equation}
When two inputs $\varrho_{+}$ and $\varrho_{-}$ are equiprobable,
the relative error is defined by
$R_{NL}=\bigl(\sin\delta'_{+}+\sin\delta'_{-}\bigr)\big/\sin\Delta^{(L)}_{\pm}$ 
\cite{rast1,rast3}. Meaning $p_{\pm}=1/2$, it can be rewritten as
\begin{equation}
R_{NL}=\sum\nolimits_{j=\pm}\>p_j
\>\frac{\sin\delta'_j}{(1/2)\sin\Delta^{(L)}_{\pm}}
\equiv\sum\nolimits_{j=\pm}\>p_j\>
\frac{d(\varrho'_j,\rho_j^{\otimes L})}{(1/2)d(\rho_{+}^{\otimes L},\rho_{-}^{\otimes L})}
\ . \label{bd6}
\end{equation}
The right-hand side of Eq. (\ref{bd6}) is quite relevant to the
case of arbitrary prior probabilities. Since the distance
$d(\varrho'_j,\rho_j^{\otimes L})$ estimates the difference
between two probability distributions (see Eq. (\ref{pr11})), a
reliable identification of original input via measurement over
clones may be provided only when
$$
d(\varrho'_{+},\rho_{+}^{\otimes L})+d(\varrho'_{-},\rho_{-}^{\otimes L})\ll
d(\rho_{+}^{\otimes L},\rho_{-}^{\otimes L}) \ .
$$
So we see a reason for using a ratio of
$d(\varrho'_j,\rho_j^{\otimes L})$ just to the half of
$d(\rho_{+}^{\otimes L},\rho_{-}^{\otimes L})$. In addition, this
choice implies that the tight lower bound on relative error
generally recovers the range $[0;1]$.

We shall now extend the notion of relative error for the set
${\mathfrak{S}}=\{\rho_1,\ldots,\rho_m\}$ with $m>2$, when the
number of different pairs is equal to $m(m-1)/2$. The probability
of taking the pair $\pi_{jk}=\{\rho_j,\rho_k\}$ is equal to
\begin{equation}
q_{jk}=p_jp_k   \left(\sum\nolimits_{1\leq j<k\leq m}p_jp_k\right)^{-1}
\ , \label{qujk}
\end{equation}
where $\sum\nolimits_{1\leq j<k\leq m}q_{jk}=1$. We clearly have
$p_j=1/m$ and $q_{jk}=2/\bigl(m(m-1)\bigr)$ for the set of $m$
equiprobable states. To each pair $\pi_{jk}$ assign the quantity
\begin{equation}
R_{NL}(\pi_{jk})=
2\left\{p_j{\>}d(\varrho'_j,\rho_j^{\otimes L})+p_k{\>}d(\varrho'_k,\rho_k^{\otimes L})\right\}
\big/\bigr\{(p_j+p_k){\>}d(\rho_{j}^{\otimes L},\rho_{k}^{\otimes L})\bigl\}
{\>}, \label{aveid}
\end{equation}
which takes into account that, perhaps, $p_j+p_k\neq1$. It is
natural to put the weighted average of the $m(m-1)/2$ quantities
(\ref{aveid}).

{\bf Definition 1.} {\it The relative error of $N\to L$ cloning of
the set ${\mathfrak{S}}=\{\rho_1,\ldots,\rho_m\}$ is defined by}
\begin{equation}
R_{NL}({\mathfrak{S}}):=\sum\nolimits_{1\leq j<k\leq m}
q_{jk}\>R_{NL}(\pi_{jk})
\ . \label{bd7}
\end{equation}

Let the prior probability be value of order $\epsilon\ll1$ for all
the states except $\rho_1$ and $\rho_2$. That is, we take
$p_j=O(\epsilon)$ for $j\not=1,2$, whence $q_{12}=1+O(\epsilon)$,
$q_{jk}=O(\epsilon)$ for the rest pairs. The expression
(\ref{bd7}) for relative error is simply reduced to
$R_{NL}({\mathfrak{S}})=R_{NL}(\pi_{12})+O(\epsilon)$. In the same
manner, we can find
$R_{NL}({\mathfrak{S}})=R_{NL}({\mathfrak{T}})+O(\epsilon)$, when
${\mathfrak{T}}\subset{\mathfrak{S}}$ and probabilities
$p_j=O(\epsilon)$ except for the states $\rho_j\in{\mathfrak{T}}$
solely.

We are interested in a nontrivial lower bound on the relative
error (\ref{bd7}). Our approach to obtaining the limits utilizes
triangle inequalities \cite{rast1,rast2}. Following the method, we
shall derive the angle relation from which bound on the relative
error is simply obtained. It is handy to introduce the angle
$\varkappa_{jk}\in[0;\pi/2]$ as
\begin{equation}
\varkappa_{jk}:=\arccos
\sqrt{F(\rho_j^{\otimes N},\rho_k^{\otimes N})F(\Upsilon_j,\Upsilon_k)}
\ . \label{alpha1}
\end{equation}
The laws of quantum theory impose some restrictions on acceptable
values of angles $\delta'_j$, whence nontrivial bounds for
different figures of cloning merit follow.

\section{Lower bound for two-state set}

In the case of the two-state set $\{\rho_{+},\rho_{-}\}$, the
initial state of ancilla is $\Upsilon_{+}$ or $\Upsilon_{-}$
according to the input which is $\rho_{+}$ or $\rho_{-}$. We
further assume that
\begin{equation}
F(\Upsilon_{+},\Upsilon_{-})>
F(\rho_{+}^{\otimes M},\rho_{-}^{\otimes M})
\ ,
\label{upsres}
\end{equation}
and, by the multiplicativity of fidelity,
$\cos\varkappa_{\pm}>F(\rho_{+}^{\otimes L},\rho_{-}^{\otimes
L})$. The motivation is as follows. If the inequality
(\ref{upsres}) is not satisfied then there are states sufficient
for perfect cloning \cite{rast3,rast4}. That is, there exist states
$\Upsilon_{+}$ and $\Upsilon_{-}$ such that $\rho_j^{\otimes
M}={\rm Tr}_{E}(\Upsilon_j)$. Hence we can mention a trivial bound
$R_{NL}\geq0$ only. So we presuppose that the inequality
(\ref{upsres}) is valid. As result, we have
\begin{equation}
\Delta^{(N)}_{\pm}\leq\varkappa_{\pm}<\Delta^{(L)}_{\pm} \ .
\label{alpdel}
\end{equation}
With no loos of generality, we assume that $p_{+}\geq p_{-}$.

{\bf Theorem 2} {\it The relative error $R_{NL}$ of cloning of the
set $\{\rho_{+},\rho_{-}\}$ satisfies}
\begin{equation}
R_{NL}\geq 2{\,}p_{-} \sin(\Delta^{(L)}_{\pm}-\varkappa_{\pm})\>\sin^{-1}\Delta^{(L)}_{\pm}
\ . \label{theor3}
\end{equation}

{\bf Proof}
Applying the inequality (\ref{pr10}) twice, we obtain
\begin{equation}
\Delta^{(L)}_{\pm}\leq\Delta(\rho_{+}^{\otimes L},\varrho'_{+})
+\Delta(\rho_{-}^{\otimes L},\varrho'_{+})\leq
\delta'_{+}+\delta'_{-}+ \Delta(\varrho'_{+},\varrho'_{-})
\ . \label{tilde1}
\end{equation}
Recall that the fidelity function is multiplicative, preserved by
unitary evolution  and non-decreasing under the operation of
partial trace \cite{barnum,jozsa2}. So we obtain
$$
F(\rho_{+}^{\otimes N},\rho_{-}^{\otimes N}) {\>} F(\Upsilon_{+},\Upsilon_{-})\leq
F(\varrho'_{+},\varrho'_{-}) \ ,
$$
whence $\Delta(\varrho'_{+},\varrho'_{-})\leq\varkappa_{\pm}$.
Combining this with Eq. (\ref{tilde1}) provides
\begin{equation}
\delta'_{+}+\delta'_{-} \geq
\Delta^{(L)}_{\pm}-\varkappa_{\pm}
\ . \label{star1}
\end{equation}
Consider the function $g(\delta'_{+},\delta'_{-}):=p_{+}\sin\delta'_{+}+p_{-}\sin\delta'_{-}$
to be minimized. We want to minimize $g(\delta'_{+},\delta'_{-})$ under the
constraint (\ref{star1}), $0\leq\delta'_{+}\leq\pi/2$ and
$0\leq\delta'_{-}\leq\pi/2$. This task is solved in Appendix
A. By substitutions, we then have
\begin{equation}
\min g(\delta'_{+},\delta'_{-})=p_{-} \sin(\Delta^{(L)}_{\pm}-\varkappa_{\pm})
\end{equation}
and further the statement of Theorem 2. $\square$

For equiprobable states, the bound (\ref{theor3}) is reduced to
the lower bound deduced in Ref. \cite{rast3}. In terms of
$f=\sqrt{F}(\rho_{+},\rho_{-})$ and
$\phi=\sqrt{F}(\Upsilon_{+},\Upsilon_{-})$, we rewrite
(\ref{upsres}) as $\phi>f^M$. By $\cos\Delta^{(L)}_{\pm}=f^L$ and
$\cos\varkappa_{\pm}=f^M\phi$, the bound (\ref{theor3}) becomes
\begin{equation}
R_{NL}\geq 2\>p_{-}\Bigl\{
f^M\phi- f^L\sqrt{\bigl(1-f^{2M}\phi^2\bigr){\big/}\bigl(1-f^{2L}\bigr)}
{\,}\Bigr\}
\ . \label{theor33}
\end{equation}
At fixed $f$ and $\phi$, the right-hand side of Eq.
(\ref{theor33}) is an increasing function of probability $p_{-}$.
That is, it decreases as the prior probabilities differ. This is
analog of that the upper bound on the global fidelity increases in
such a situation \cite{rast4}. We are rather interested in
dependence of the bound on $\phi$. This parameter marks a top
amount of an {\it a priori} information, which can initially be
laid in the ancilla. The more a value of $\phi$, the less this
amount. The angle $\varkappa_{\pm}$ is a decreasing function of
$\phi$. In the range (\ref{alpdel}), the lower bound by Theorem 3
is a decreasing function of $\varkappa_{\pm}$. So the right-hand
side of (\ref{theor33}) increases as the marker $\phi$ of
additional information increases. For $\phi=f^{M}$ the perfect
cloning can be reached \cite{rast3,rast4}. In line with this fact,
we have $\varkappa_{\pm}=\Delta^{(L)}_{\pm}$ and the vanishing
bound on $R_{NL}$. On the contrary, in the usual cloning there is
no {\it a priori} information, i.e. $\Upsilon_{+}=\Upsilon_{-}$
and $\phi=1$. Then the bound by Theorem 2 reaches its maximum as a
function of $\phi$. The above points reproduce the observations of
Ref. \cite{rast3} in more general setting.

If $N\to\infty$ at fixed $M$ then the right-hand side of the
inequality (\ref{theor33}) goes to zero. This is natural because
infinite number $N$ of originals can provide almost perfect
cloning. If $M\to\infty$ at fixed $N$ then the right-hand side of
(\ref{theor33}) recovers the value $2{\,}p_{-}f^N\phi$. In the
standard cloning of equiprobable states ($\phi=1$, $p_{-}=1/2$),
this value can be arbitrarily close to 1, since $\sup\{2{\,}p_{-}
f^N \phi:{\,}0\leq f^M<\phi\leq 1\}=1$. It is not insignificant
that the value $2{\,}p_{-}f^N\phi$ gives the minimal size of
probability of inconclusive answer for unambiguous discrimination
at $p_{-}=1/2$. Namely, the success discrimination of the
equiprobable pure states
$|\Omega_{\pm}\rangle=|\psi_{\pm}\rangle^{\otimes
N}\otimes|\theta_{\pm}\rangle$ occurs with the optimal
probability
$\bigl(1-|\langle\Omega_{+}|\Omega_{-}\rangle|\bigr)$ \cite{ivan1,peres1}. The value
$2{\,}p_{-}f^N\phi$ is obtained for
$|\langle\psi_{+}|\psi_{-}\rangle|=f$ and
$|\langle\theta_{+}|\theta_{-}\rangle|=\phi$. Note that the upper
bound on the global fidelity in the limit $M\to\infty$ at fixed
$N$ goes to well-known Helstrom bound \cite{chefles2,rast4}. It is
the probability of correctly distinguishing between two pure
states $|\Omega_{\pm}\rangle$ by the optimal
strategy \cite{helstrom}.

\section{Lower bound for multi-state set}

We now obtain a lower bound on the relative error of cloning of
the set ${\mathfrak{S}}=\{\rho_1,\ldots,\rho_m\}$. As before, the
prior probabilities are arbitrary and constrained only by the
normalization condition. Like (\ref{alpdel}), we have the
acceptable range
\begin{equation}
\Delta^{(N)}_{jk}\leq\varkappa_{jk}<\Delta^{(L)}_{jk} \ .
\label{alpdel2}
\end{equation}
According to Theorem 2, each term of sum in the right-hand side of
(\ref{bd7}) obeys
\begin{equation}
R_{NL}(\pi_{jk})\geq 2\min\{p_j,p_k\}\sin(\Delta^{(L)}_{jk}-\varkappa_{jk})
\left((p_j+p_k){\,}\sin\Delta^{(L)}_{jk}\right)^{-1}
\ . \label{rjk1}
\end{equation}
Hence the desired bound is established as follows.

{\bf Theorem 3} {\it The relative error of $N\to L$ cloning of the
set ${\mathfrak{S}}=\{\rho_1,\ldots,\rho_m\}$ satisfies}
\begin{equation}
R_{NL}({\mathfrak{S}})\geq\sum_{1\leq j<k\leq m}
2{\>}q_{jk}{\,}\frac{\min\{p_j,p_k\}}{p_j+p_k}{\ }\frac{\sin(\Delta^{(L)}_{jk}-\varkappa_{jk})}{\sin\Delta^{(L)}_{jk}}
\ . \label{theor4}
\end{equation}

As a straightforward extension, the bound (\ref{theor4}) succeeds
many features of the bound (\ref{theor3}). If two probabilities,
say, $p_1$ and $p_2$ are variable and the rest of parameters is
fixed, then the bound (\ref{theor4}) decreases as these
probabilities differ. If some one probability is close to 1 and
other probabilities are small, then the bound is close to zero.
This behavior is expected, because single known state can be
cloned perfectly. For equal {\it a priori} probabilities
$p_j=1/m$, the bound by Theorem 3 becomes
\begin{equation}
R_{NL}\geq\frac{2}{m(m-1)}\sum_{1\leq j<k\leq m}
\bigl( \cos\varkappa_{jk}-\sin\varkappa_{jk}\cot\Delta^{(L)}_{jk} \bigr)
\ . \label{prieq}
\end{equation}
It is natural that both the bounds given by (\ref{theor4}) and
(\ref{prieq}) decrease as $\varkappa_{jk}$ increases.  Indeed, the
parameter $\varkappa_{jk}$ characterizes an amount of prior
information. If the upper limit of Eq. (\ref{alpdel2}) is
saturated for some pair $\pi_{jk}$ then corresponding summands in
the right-hand sides of Eqs. (\ref{theor4}) and (\ref{prieq})
vanish. This is the case of potentially perfect cloning. On the
whole, these conclusions on a role of {\it a priori} information
in the ancilla add to the stronger no-cloning theorem.

A question is, whether the lower bounds (\ref{theor3}) and
(\ref{theor4}) can be reached? In general, it is not the case,
though the bound by Theorem 3 is least for two pure states. The
quantum circuit for optimal cloning will be built in the next
section. The subject matter changes for $m>2$. From the viewpoint
of minimization the bound of Theorem 3 is approximate. As reasons
of Appendix A show, saturating the inequality (\ref{rjk1}) holds
if and only if $\delta'_{j}=0$,
$\delta'_k=\Delta^{(L)}_{jk}-\varkappa_{jk}$ for $p_j\geq p_k$
(for $p_j\leq p_k$ the angles $\delta'_{j}$ and $\delta'_{k}$
should be swapped in the two equalities). These two equalities per
each of $m(m-1)/2$ pairs totally give $m(m-1)$ conditions. For
saturating Eq. (\ref{theor4}), $m$ variables $\delta'_{j}$ must
satisfy all these $m(m-1)$ conditions. Except for some special
cases, this is not possible. Thus, the presented limit is somewhat
rough.

More rigorous way may be as follows. Similar to (\ref{star1}), we
have arrived at the $m(m-1)/2$ inequalities of a kind
$\delta'_{j}+\delta'_{k}\geq\Delta^{(L)}_{jk}-\varkappa_{jk}$.
Together with the $m$ conditions $0\leq\delta'_j\leq\pi/2$, these
relations specify some simplex in $m$-dimensional real space. The
relative error (\ref{bd7}) can be rewritten in the form
\begin{equation}
R_{NL}({\mathfrak{S}})=2\sum\nolimits_{j=1}^{m}
r_jp_j\sin\delta'_j
\ , \label{bdel7}
\end{equation}
where
$r_j=\sum\nolimits_{k\not=j}q_{jk}\bigl((p_j+p_k){\,}\sin\Delta^{(L)}_{jk}\bigr)^{-1}$.
The task is to minimize the function (\ref{bdel7}) in the above
simplex. So we come across a difficult problem of nonlinear
programming (the simple case $m=2$ of this problem is considered
in Appendix A). For $0\leq\delta'_j\leq\pi/2$, the minimized
function is concave. So the problem of minimization is reduced to
finding extremal points of the simplex. If the values of
parameters are prescribed, the wanted minimum can be found
numerically. At the same time, it is complicated to obtain an
explicit formula for general case. But even if we should find it,
we still would not have a complete solution to the problem of
mixed-state cloning. Indeed, it is not necessary that bound given
by such a formula be least. So we have restricted our
consideration to obtaining of the bound by Theorem 3. Rough though
this bound is, it has straightforward form and allows to estimate
how a merit of state-dependent cloning is limited.

\section{Comparison of different criteria}

We shall now expose the relative error in comparison with other
optimality criteria. For the sake of simplicity, we restrict to
the $N\to L$ cloning of two equiprobable pure states
$|\psi_{\pm}\rangle$ without {\it a priori} information in the
ancilla. How able to good cloning is the pair? This question is
central to applications of quantum cloning. In principle, we may
assume both the deterministic cloning and probabilistic
cloning \cite{duan2}. A merit of deterministic cloning may be
viewed with respect to the global fidelity, the absolute error and
the relative error. For equiprobable inputs, the global fidelity
is expressed by
$F_{NL}={2}^{-1}\cos^2\delta'_{+}+{2}^{-1}\cos^2\delta'_{-}$ \cite{chefles2,rast2}.
Hillery and Bu\v{z}ek \cite{hillery2} used the measure
$A_{NL}={2}^{-1}\sin\delta'_{+}+{2}^{-1}\sin\delta'_{-}$. This
measure will be referred to as {\it absolute error} \cite{rast1}.
The relative error is defined by Eq. (\ref{bd6}). In probabilistic
cloning, the exact clone of an input successfully generated with
the maximal probability \cite{duan2,chefles1}
\begin{equation}
\max P_{NL}=\bigl(1-f^N\bigr)\big/\bigl(1-f^L\bigr)
\ , \label{prexp}
\end{equation}
where $f$ denote the overlap $|\langle\psi_{+}|\psi_{-}\rangle|$
between states $|\psi_{\pm}\rangle$. As it is shown in
Refs. \cite{chefles2,rast3}, the maximum of the global fidelity
is equal to
\begin{equation}
\max F_{NL} =\frac{1}{2}\left(1+f^{L+N}+\sqrt{(1-f^{2N})(1-f^{2L})}{\,}\right)
\ . \label{opfid}
\end{equation}
According to (\ref{theor33}), the minimum of the relative error is reduced to
\begin{equation}
\min R_{NL}= f^N- f^L\left(1-f^{2N}\right)^{1/2}\left(1-f^{2L}\right)^{-1/2}
\label{theor333}
\end{equation}
for $\phi=1$ and $p_{\pm}=1/2$. For the absolute error we have \cite{rast1}
\begin{equation}
\min A_{NL}= f^N\sqrt{1-f^{2L}}- f^L\sqrt{1-f^{2N}}
\ . \label{theor303}
\end{equation}
Let us consider the two cases: (i) the states are
$|\psi_{\pm}\rangle$ are almost orthogonal, i.e. $f=\varepsilon\ll
1$; (ii) the states are $|\psi_{\pm}\rangle$ are almost identical,
i.e. $f=1-\epsilon$ with $\epsilon\ll 1$. A behaviour of each of
the criteria is shown in Table 1 ($N<L$).

\begin{table}
\caption{An asymptotic behaviour of the four criteria.}
\begin{ruledtabular}
\begin{tabular}{ccc}
Figure of merit & (i) $f=\varepsilon\ll1$ & (ii) $f=1-\epsilon$ ($\epsilon\ll1$) \\
\hline
 & & \\
$\max F_{NL}=$ & $1-\varepsilon^{2N}{\big/}{\,}4+\cdots$ & $1-(\sqrt{L}-\sqrt{N})^2\epsilon{\big/}{\,}2+\cdots$  \\
 & & \\
$\min A_{NL}=$ & $\varepsilon^{N}+\cdots$ & $(\sqrt{2L}-\sqrt{2N}){\,}\epsilon^{1/2}+\cdots$ \\
 & & \\
$\min R_{NL}=$ & $\varepsilon^{N}+\cdots$ & $1-\sqrt{N/L}+(\sqrt{LN}-N)\epsilon+\cdots$  \\
 & & \\
$\max P_{NL}=$ & $1-\varepsilon^{N}+\cdots$ & $N{\big/}L-N(L-N)\epsilon{\big/}(2L)+\cdots$ \\
\end{tabular}
\end{ruledtabular}
\end{table}

As it is clear from the second column, for the case (i) all the
measures endorse a good merit of both the deterministic and
probabilistic cloning. In effect, the optimum of global fidelity
is close to one, the optimum of absolute and relative error is
close to zero. The probability of success is close to one. It is
natural because orthogonal states can perfectly be cloned. The
principal distinction of the relative error is revealed in the
case (ii). It seems offhand that two almost identical states can
be cloned very well. Both the global fidelity and absolute error
approve the conclusion ($\max F_{NL}\approx1$ and $\min
A_{NL}\approx0$). It would be rash to accept this. In effect, the
optimal probability $\max P_{NL}$ is generally not close to one.
The first term $N/L$ is almost one only if the number $M=L-N$ of
actual clone is negligible in comparison with the number $N$ of
originals. In line with this, the optimum of relative error is
close to zero for $N/L\approx1$. But the probability is close to
zero and the relative error is close to one when the number
$M=L-N$ of actual clone is large. We see that both the global
fidelity and absolute error lose sight of the important aspect of
deterministic cloning. Even for the primary $1\to2$ cloning, we
have $\max P_{NL}=1/2-\epsilon/4+\cdots$ and $\min
R_{NL}=1-2^{-1/2}+O(\epsilon)\approx 0.3+O(\epsilon)$, that is
both the probabilistic and deterministic strategies are restricted
enough. In contrast with the global fidelity and the absolute
error, for the case (ii) a behaviour of relative error is
crucially dependent on numbers $N$ and $L$. Similar to the optimal
probability of success, the criterion of relative error emphasizes
that any cloning is not isolated stage in quantum information
processing. As a rule, the outputs of cloning machine are subjects
of further operations, say, a discrimination. For example, in the
cryptographic B92 scheme Alice encodes the bits into two
non-orthogonal pure states \cite{bennett3}. So Bob can apply the
unambiguous discrimination \cite{palma}. But the closer used states
are to each other the larger number of discarded bits is in the
total sequence. On the other hand, a sufficiently great closeness
of the used states will prevent the eavesdropping. Unlike both the
global fidelity and absolute error, the notion of relative error
allows to take such aspects into account.

\section{Circuit for optimal cloning of pure states}

We shall now build quantum circuits for the optimal relative-error
cloning of two pure states $|\psi_{\pm}\rangle$ with arbitrary
prior probabilities $p_{\pm}$ ($p_{+}\geq p_{-}$). {\it A priori}
information about actually input state is contained in the state
of ancilla which is either $|\theta_{+}\rangle$ or
$|\theta_{-}\rangle$. Without loss of generality, we take the
product $\langle\psi_{+}|\psi_{-}\rangle$ to be positive real.
These states are parametrized as
$|\psi_{\pm}\rangle=\cos\alpha_0|0\rangle\pm\sin\alpha_0|1\rangle\equiv|\varphi_{\pm}(\alpha_0)\rangle$,
$|\theta_{\pm}\rangle=\cos\theta|0\rangle\pm\sin\theta|1\rangle\equiv|\varphi_{\pm}(\theta)\rangle$.
The overlap is $\langle\psi_{+}|\psi_{-}\rangle=\cos 2\alpha_0$
with $\alpha_0\in(0;\pi/4)$. So, we have the register of $(L+1)$
qubits, where $M$ qubits are initially in the blank state
$|0\rangle$, $N$ qubits are in the state to be cloned, and one
qubit is ancillary. Our aim is to transform these states according
to the specification.

The strategy is an extension of the known one \cite{chefles2} and
uses the {\it distinguishability transfer gate} (see Appendix B).
First, the information about the input $N$ originals is
transferred into one qubit. We mark the ancillary qubit by ''0'',
the $N$ original qubits by ''$1,\ldots,N$'', and the $M$
additional qubits by ''$N+1,\ldots,L$''. The just left gate acts
on the qubits $(N-1)$ and $N$ as
\begin{equation}
\D_N(\alpha_0,\alpha_0)\>|\varphi_{\pm}(\alpha_0)\rangle_{N-1}\otimes|\varphi_{\pm}(\alpha_0)\rangle_{N}
=|\varphi_{\pm}(\alpha_1)\rangle_{N-1}\otimes|0\rangle_{N}
\>, \label{gdn}
\end{equation}
where $\cos2\alpha_1=(\cos2\alpha_0)^2$. Then an operation is
applied to qubits $(N-2)$ and $(N-1)$, and so on. In the first
stage, the gate $\D_j$ transfers the distinguishability from $j$th
qubit to $(j-1)$th ($j$ runs from $N$ to $2$), i.e.
\begin{equation}
\D_j(\alpha_0,\alpha_{N-j})\>|\varphi_{\pm}(\alpha_0)\rangle_{j-1}\otimes|\varphi_{\pm}(\alpha_{N-j})\rangle_{j}
=|\varphi_{\pm}(\alpha_{N-j+1})\rangle_{j-1}\otimes|0\rangle_{j}
\ , \label{gdl}
\end{equation}
where $\cos2\alpha_0\cos2\alpha_{N-j}=\cos2\alpha_{N-j+1}$,
$\cos2\alpha_{N-1}=(\cos2\alpha_0)^{N}$. Within the first stage,
the state changes as
\begin{equation}
\left\{\sideset{_{j=2}}{_N}\prod
\D_j(\alpha_0,\alpha_{N-j})\right\}
|\varphi_{\pm}(\alpha_0)\rangle^{\otimes N}=
|\varphi_{\pm}(\alpha_{N-1})\rangle_{1}\otimes|0\rangle^{\otimes(N-1)}
\ , \label{stag1}
\end{equation}
where the gates ${\D}_j$ are put from right to left with
decreasing $j$. This part transfers a total distinguishability of
the $N$ originals $|\varphi_{\pm}(\alpha_0)\rangle$ into the
one-qubit state $|\varphi_{\pm}(\alpha_{N-1})\rangle$. An example
for $3\to5$ cloning is shown on Fig. 1.

\begin{figure}
\centering
\includegraphics{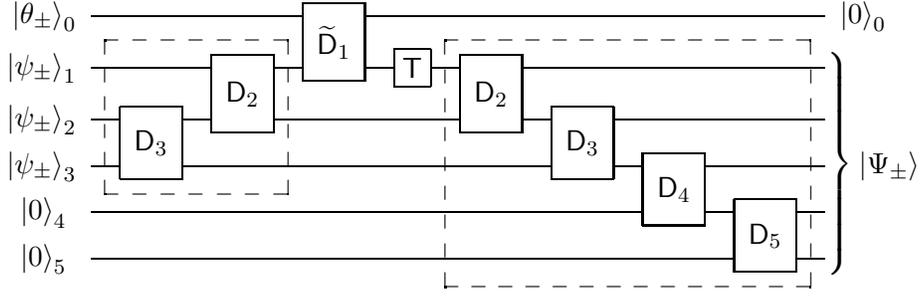}
\caption{The
circuit for optimal $3\to5$ cloning with the {\it a priori}
information. For brevity, the gate $\D_j(\alpha_0,\alpha_{N-j})$
in the left box and the gate $\D_k(\alpha_0,\alpha_{L-k})$ the
right box are both denoted as $\D_k$.}
\end{figure}

For using an {\it a priori} information, we now include the {\it
turned} gate $\widetilde{\D}_1$. This gate transfers
distinguishability of ancilla's states $|\theta_{\pm}\rangle$ to
those of qubit 1, namely
\begin{equation}
\widetilde{\D}_1(\theta,\alpha_{N-1})\>|\varphi_{\pm}(\theta)\rangle_0\otimes|\varphi_{\pm}(\alpha_{N-1})\rangle_{1}
=|0\rangle_0\otimes|\varphi_{\pm}(\theta_1)\rangle_{1}
\ . \label{tildd}
\end{equation}
After the action of gate $\widetilde{\D}_1$, the ancilla contains
no information about distinguishability. All the
distinguishability of inputs are now concentrated on two possible
states $|\varphi_{\pm}(\theta_1)\rangle$ of qubit 1, where
$\cos2\theta_1=\cos2\theta\cos2\alpha_{N-1}$. Now the scheme acts
on the qubit 1 by the unitary operator $\T$ specified as
\begin{equation}
\T{\>}|\varphi_{\pm}(\theta_1)\rangle=\mu_{\pm}\>|\varphi_{+}(\alpha_{L-1})\rangle
+\nu_{\pm}\>|\varphi_{-}(\alpha_{L-1})\rangle
\ . \label{datet}
\end{equation}
The values of angle $\alpha_{L-1}$ and complex numbers $\mu_{\pm}$
and $\nu_{\pm}$ will be found below. So the second stage results
in the final state
\begin{equation}
|\Phi_{\pm}\rangle=\mu_{\pm}\>|\varphi_{+}(\alpha_{L-1})\rangle_1\otimes|0\rangle^{\otimes(L-1)}
+\nu_{\pm}\>|\varphi_{-}(\alpha_{L-1})\rangle_1\otimes|0\rangle^{\otimes(L-1)}
\label{secfin}
\end{equation}
of the $L$ qubits. In Fig. 1, the gates $\widetilde{\D}_1$ and
$\T$ between the dash boxes perform the second stage. Its
structure is independent of numbers $N$ and $L$. Note that this
stage and an {\it a priori} information are not considered in
Ref.\cite{chefles2}.

We put two linear combinations of the ideal outputs
$|\psi_{+}\rangle^{\otimes L}$ and $|\psi_{-}\rangle^{\otimes L}$
as
\begin{equation}
|\Psi_{\pm}\rangle:=\mu_{\pm}\>|\varphi_{+}(\alpha_0)\rangle^{\otimes L}
+\nu_{\pm}\>|\varphi_{-}(\alpha_0)\rangle^{\otimes L} \ .
\label{tildd3}
\end{equation}
The final stage of cloning is posed as
$|\Phi_{\pm}\rangle\longmapsto|\Psi_{\pm}\rangle$. Let us continue the sequence
$\{\alpha_0,\alpha_1,\ldots,\alpha_{N-1}\}$ with respect to
the above recurrence, that is
\begin{equation}
\cos2\alpha_{L-k+1}=\cos2\alpha_0\cos2\alpha_{L-k}
\ , \label{csgdll}
\end{equation}
where $\alpha_{L-k}\in[0;\pi/4]$. Hence we obtain
$\cos2\alpha_{L-1}=(\cos2\alpha_0)^L$. Due to the property
(\ref{dp3012}) of distinguishability transfer gate, we have
\begin{equation}
\D_k(\alpha_0,\alpha_{L-k})\>|\varphi_{\pm}(\alpha_{L-k+1})\rangle_{k-1}\otimes|0\rangle_{k}
=|\varphi_{\pm}(\alpha_0)\rangle_{k-1}\otimes|\varphi_{\pm}(\alpha_{L-k})\rangle_{k}
\ . \label{gdll}
\end{equation}
In the third stage, the label $k$ in
(\ref{gdll}) runs from $k=2$ to $k=L$. So, the gate
$\D_2(\alpha_0,\alpha_{L-2})$ acts on the qubits 1 and 2, the gate
$\D_3(\alpha_0,\alpha_{L-3})$ acts on the qubits 2 and 3, and so on.
The total action is described by
\begin{equation}
\left\{\sideset{_L}{_{k=2}}\prod\nolimits
\D_k(\alpha_0,\alpha_{N-k})\right\}
|\varphi_{+}(\alpha_{L-1})\rangle_1\otimes|0\rangle^{\otimes(L-1)}
=|\varphi_{\pm}(\alpha_0)\rangle^{\otimes L}
\ , \label{stag3}
\end{equation}
where the gates ${\D}_k$ are put from right to left with
increasing $k$. In (\ref{stag3}), the accumulated
distinguishability is distributed among the $L$ qubits of
interest. On Fig. 1, the four gates $\D_2$, $\D_3$, $\D_4$ and
$\D_5$ of the third stage are grouped in the right dash box. Using
the linearity, we see that
$|\Phi_{\pm}\rangle\longmapsto|\Psi_{\pm}\rangle$ too. Due to
$\langle\Phi_{+}|\Phi_{-}\rangle=\langle\Psi_{+}|\Psi_{-}\rangle$,
\begin{equation}
(\mu_{+}^{*}\nu_{-}+\nu_{+}^{*}\mu_{-})\cos2\alpha_{L-1}=
(\mu_{+}^{*}\nu_{-}+\nu_{+}^{*}\mu_{-})(\cos2\alpha_{0})^L
\ , \label{unit1}
\end{equation}
that is actually correct. Specifying concrete values of $\mu_{\pm}$
and $\nu_{\pm}$ and herewith the single-qubit gate $\T$ in Eq.
(\ref{datet}), we can optimize either the relative error
or the global fidelity. In each case, we superpose the
${\rm{span}}\{|\varphi_{+}(\theta_1)\rangle,|\varphi_{-}(\theta_1)\rangle\}$
onto the
${\rm{span}}\{|\varphi_{+}(\alpha_{L-1})\rangle,|\varphi_{-}(\alpha_{L-1})\rangle\}$.
Then after the second stage the $L$ qubits of interest lie in the
states $|\Phi_{\pm}\rangle$. For the optimality with respect to
the relative error, we demand that $\delta'_{+}=0$, whence we get
$|\Psi_{+}\rangle=|\varphi_{+}(\alpha_0)\rangle^{\otimes L}$ and
$\mu_{+}=1$, $\nu_{+}=0$ from (\ref{tildd3}). The angle between
$|\varphi_{+}(\theta_1)\rangle$ and
$|\varphi_{-}(\theta_1)\rangle$ is equal to $2\theta_1$, the angle
between $|\varphi_{+}(\alpha_{L-1})\rangle$ and
$|\varphi_{-}(\alpha_{L-1})\rangle$ is equal to
$2\alpha_{L-1}>2\theta_1$. Because unitary transformations
preserve angles, the angle between
$\T|\varphi_{-}(\theta_1)\rangle$ and
$|\varphi_{-}(\alpha_{L-1})\rangle$ is equal to
$\delta'_{-}=2\alpha_{L-1}-2\theta_1$ Within the third stage, the
state
$|\Phi_{-}\rangle=\T|\varphi_{-}(\theta_1)\rangle_1\otimes|0\rangle^{\otimes(L-1)}$
maps to $|\Psi_{-}\rangle$. By definition, the value $\delta'_{-}$
is angle between $|\Psi_{-}\rangle$ and
$|\varphi_{-}(\alpha_0)\rangle^{\otimes
L}=|\psi_{-}\rangle^{\otimes L}$. Since
$\cos\Delta^{(L)}_{\pm}=(\cos2\alpha_0)^L=\cos2{\,}\alpha_{L-1}$
and
$\cos\varkappa_{\pm}=(\cos2\alpha_0)^N\cos2\theta=\cos2\theta_1$,
we find the needed value
$\delta'_{-}=\Delta^{(L)}_{\pm}-\varkappa_{\pm}$. Thus, the
inequality (\ref{theor3}) is saturated too, and the built scheme
is really optimal with respect to the relative error.

Note that $\mu_{-}$ and $\nu_{-}$ are found as
$\mu_{-}=\sin(\Delta^{(L)}_{\pm}-\varkappa_{\pm})/\sin\Delta^{(L)}_{\pm}$
and $\nu_{-}=\sin\varkappa_{\pm}/\sin\Delta^{(L)}_{\pm}$. But the
described geometrical picture is quite sufficient for all the
purposes. In the same manner, the optimization of cloning with
respect to the global fidelity would be considered. As result, the
generalization of the deterministic cloner of
Ref. \cite{chefles2} to prior ancillary information can be
obtained.

\section{Conclusion}

We have analyzed a new optimality criterion for the
state-dependent cloning of several states with arbitrary prior
probabilities and an ancillary information. The notion of the
relative error has been extended to the general cloning scenario.
The lower bounds on the relative error have been obtained for both
the two-state and multi-state cases. The attainability of the
derived bounds has been discussed. The quantum circuit for optimal
cloning of two pure states with respect to the relative error has
been built. Our approach is based on the simple geometrical
description, which generally clarifies origins of a bound for one
or another figure of merit. In principle, the described scheme
allows to develop cloning circuit that is optimal with respect to
any non-local figure of merit. The scenario with an {\it a priori}
information in the ancilla was inspired by the stronger no-cloning
theorem. The obtained conclusions on a possible merit of the
cloning contribute to this subject. Unequal prior probabilities of
inputs are usual in communication systems. The examination of
mixed-state cloning is needed because all the real devices are
inevitably exposed to noise. Analysis with respect to the relative
error may have potential applications to the problem of
eavesdropping in quantum cryptography.

\section*{Appendix A. Lemma}

Let us consider the function $f(x,y)=p\sin x+q\sin y$, where
positive $p$ and $q$ obey $\,p+q=1\,$. Let $\,a\in[0;\pi/2]\,$ be
a fixed parameter. The range of variables is stated by conditions
$\,x+y\geq a\,$, $\,0\leq x\leq\pi/2\,$ and $\,0\leq
y\leq\pi/2\,$. This domain $D$ is a square whose left-lower corner
is cut off by line $\,x+y=a\,$.

{\bf Lemma 4} {\it The global minimum of the function $f(x,y)$ in the
domain $D$ is equal to} $f_{min}=\min\{p,q\}\sin a$.

{\bf Proof} Inside of the domain $D$, we have $\,\partial
f/\partial x\not=0\,$ and $\,\partial f/\partial y\not=0\,$. So
the extreme values are reached on the boundary $\partial D$.
Consider those segments that are parallel to either axis $x$ or
axis $y$. The minimum value on these segments is equal to either
$f(a,0)=p\sin a$ or $f(0,a)=q\sin a$, i.e. $\min\{p,q\}\sin a$.

On the segment $\,x+y=a$, we put $x=a/2+t$ and $y=a/2-t$ with
$t\in[-a/2;a/2]$, whence $f(x,y)=\sin(a/2)\cos
t+(p-q)\cos(a/2)\sin t$. By calculus, we obtain the extreme value
$(p^2-2p\>q\cos a+q^2)^{1/2}$ for $t\in[-a/2;a/2]$. This value is
not less than both the $p\sin a$ and $q\sin a$. $\square$

\section*{Appendix B. Distinguishability transfer gate}

By this operation, a distiguishability of the possible states of
second qubit is translated to those of the first. It is convenient
to introduce a family of states
$|\varphi_{\pm}(\alpha)\rangle:=\cos\alpha|0\rangle\pm\sin\alpha|1\rangle$
with the inner product
$\langle\varphi_{+}(\alpha)|\varphi_{-}(\alpha)\rangle=\cos
2\alpha$, where $\alpha\in[0;\pi/4]$. As is well-known, one- and
two-qubit gates are sufficient to implement universal computation.
In the context of cloning, the writers of Ref. \cite{chefles2}
note that only one type of pair-wise interaction is needed. The
{\it distiguishability transfer gate} is described
by \cite{chefles2}
\begin{eqnarray}
\D(\alpha,\beta)\>|\varphi_{\pm}(\alpha)\rangle_{1}\otimes|\varphi_{\pm}(\beta)\rangle_{2}
= |\varphi_{\pm}(\gamma)\rangle_{1}\otimes|0\rangle_{2}
\ , \label{dp1230} \\
\D(\alpha,\beta)\>|\varphi_{\pm}(\gamma)\rangle_{1}\otimes|0\rangle_{2}
=|\varphi_{\pm}(\alpha)\rangle_{1}\otimes|\varphi_{\pm}(\beta)\rangle_{2}
\ , \label{dp3012}
\end{eqnarray}
where by the unitarity $\cos 2\alpha\cos 2\beta=\cos 2\gamma$. It
follows from Eqs. (\ref{dp1230}) and (\ref{dp3012}) that the
operation $\D$ is Hermitian \cite{chefles2}. The action of
distiguishability transfer gate on two-qubit register is shown on
Figure 2. The corresponding circuit of $CNOT$ elements and
one-qubit operations is given in Ref. \cite{chefles2}.

\begin{figure}
\centering
\includegraphics{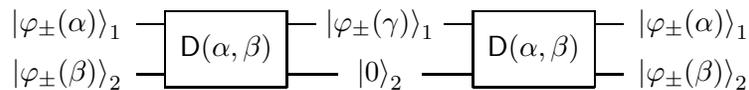}
\caption{The action of distiguishability transfer gate.}
\end{figure}


\begin{thebibliography}{55}

\bibitem{hayashi}%------------------------------------------------
M.~Hayashi, {\it Quantum Information: An Introduction} (Springer, Berlin, 2006).

\bibitem{gisin}%----------------------------------------------
N. Gisin {\it et al.}, {\it Rev. Mod. Phys.} {\bf 74} (2002) 145--195.

\bibitem{vandam}%-----------------------------------------------
A. M. Childs and W. van Dam, {\it Rev. Mod. Phys.} {\bf 82} (2009) 1--52.

\bibitem{palma}%----------------------------------------------
A. K. Ekert {\it et al.}, {\it Phys. Rev.} A {\bf 50} (1994) 1047--1056.

\bibitem{pirandola}%----------------------------------------------
S. Pirandola, {\it Int. J. Quantum Inf.} {\bf 6} (2008)  765--771.

\bibitem{wootters}%---------------------------------------------
W. K. Wootters and W. Zurek,  {\it Nature} {\bf 299} (1982) 802--803.

\bibitem{dieks}%-----------------------------------------------
D. Dieks, {\it Phys. Lett.} A {\bf 92} (1982) 271--272.

\bibitem{barnum}%-----------------------------------------------
H. Barnum {\it et al.}, {\it Phys. Rev. Lett.} {\bf 76} (1996) 2818--2821.

\bibitem{hillery1}%----------------------------------------------
V. Bu\v{z}ek and M. Hillery, {\it Phys. Rev.} A {\bf 54} (1996) 1844--1852.

\bibitem{fiurasek}%----------------------------------------------
N. J. Cerf and J. Fiur\'{a}\u{s}ek, Optical quantum cloning -- a review, quant-ph/0512172.

\bibitem{fanh}%----------------------------------------------
H. Fan, Quantum cloning machines, in {\it Quantum Computation and Information. From Theory to Experiment},
eds. H. Imai and M. Hayashi (Springer, Berlin, 2006), 63--110.

\bibitem{hillery2}%----------------------------------------------
M. Hillery and V. Bu\v{z}ek, {\it Phys. Rev.} A {\bf 56} (1997) 1212--1216.

\bibitem{bruss1}%------------------------------------------------
D. Bru{\ss} {\it et al.}, {\it Phys. Rev.} A {\bf 57} (1998) 2368--2378.

\bibitem{pati07}%-----------------------------------------------
S. Adhikari {\it et al.}, {\it Quantum Inf. Process.} {\bf 6} (2007) 197--219.

\bibitem{siomau}%-----------------------------------------------
M. Siomau and S. Fritzsche {\it Eur. Phys. J.} D {\it 57} (2010) 293--300

\bibitem{rast1}%-----------------------------------------------
A. E. Rastegin, {\it Phys. Rev.} A {\bf 66} (2002) 042304.

\bibitem{chefles2}%---------------------------------------------
A. Chefles and S. M. Barnett, {\it Phys. Rev.} A {\bf 60} (1999) 136--144.

\bibitem{guo}%-------------------------------------------------
Y. J. Han {\it et al.}, {\it Phys. Rev.} A {\bf 66} (2002) 052301.

\bibitem{kazakov}%-----------------------------------------------
A. Ya. Kazakov, {\it Int. J. Quantum Inf.} {\bf 8} (2010) 435--442.

\bibitem{bennett3}%---------------------------------------------
C. H. Bennett, {\it Phys. Rev. Lett.} {\bf 68} (1992) 3121--3124.

\bibitem{rast2}%-----------------------------------------------
A. E. Rastegin, {\it Phys. Rev.} A {\bf 67} (2003) 012305.

\bibitem{uhlmann09}%------------------------------------------------
J. A. Miszczak {\it et al.}, {\it Quantum Information} {\&}
{\it Computation} {\bf 9} (2009) 0103--0130.

\bibitem{rast3}%-----------------------------------------------
A. E. Rastegin, {\it J. Opt. B: Quantum Semiclassical Opt.} {\bf 5} (2003) S647--S650.

\bibitem{jozsa1}%-----------------------------------------------
R. Jozsa, A stronger no-cloning theorem, quant-ph/0204153.

\bibitem{rast4}%-----------------------------------------------
A. E. Rastegin, {\it Phys. Rev.} A {\bf 68} (2003) 032303.

\bibitem{maz}%----------------------------------------------
Z. H. Ma {\it et al.}, {\it Phys. Lett.} A {\bf 373} (2009) 3407--3409.

\bibitem{rast06}%-----------------------------------------------
A. E. Rastegin, Sine distance for quantum states, quant-ph/0602112.

\bibitem{uhlmann1}%-----------------------------------------------
A. Uhlmann, {\it Rep. Math. Phys.} {\bf 9} (1976) 273--279.

\bibitem{jozsa2}%-----------------------------------------------
R. Jozsa, {\it J. Mod. Optics} {\bf 41} (1994) 2315--2323.

\bibitem{rast07}%-------------------------------------------
A. E. Rastegin, {\it J. Phys. A: Math. Theor.} {\bf 40} (2007) 9533--9549.

\bibitem{rast091}%-------------------------------------------
A. E. Rastegin, {\it Quantum Inf. Process.} {\bf 9} (2010) 61--73.

\bibitem{ivan1}%-----------------------------------------------
I. D. Ivanovic, {\it Phys. Lett.} A {\bf 123} (1987) 257--259.

\bibitem{peres1}%-----------------------------------------------
A. Peres, {\it Phys. Lett.} A {\bf 128} (1988) 19.

\bibitem{helstrom}%-----------------------------------------------
C. W. Helstrom, {\it Quantum Detection and Estimation Theory}
(Academic Press, New York, 1976).

\bibitem{duan2}%-----------------------------------------------
L.-M. Duan and G.-C. Guo, {\it Phys. Lett.} A {\bf 243} (1998) 261--264.

\bibitem{chefles1}%---------------------------------------------
A. Chefles and S. M. Barnett, {\it J. Phys. A: Math. Gen.} {\bf 31} (1998) 10097--10103.



\end{thebibliography}
\end{document}